\documentstyle[12pt]{article}

\begin{document}

\hfill SOGANG-HEP 195/95

\hfill January, 1995

\vspace{1cm}

\begin{center}
{\large \bf Batalin-Tyutin Quantization of the
            Chern-Simons-Proca Theory}
\end{center}

\vspace{1cm}

\begin{center}
Ei-Byung Park, Yong-Wan Kim, Young-Jai Park, and Yongduk Kim \\
{\it Department of Physics and Basic Science Research Institute \\
Sogang University, C.P.O. Box 1142, Seoul 100-611, Korea}
\end{center}

\begin{center}
Won Tae Kim \\
{\it Center for Theoretical Physics and Department of Physics \\
Seoul National University, Seoul 151-742, Korea}
\end{center}

\vspace{1cm}

\begin{center}
{\bf ABSTRACT}
\end{center}

We quantize the Chern-Simons-Proca theory in three dimensions
by using the Batalin-Tyutin Hamiltonian method,
which systematically embeds second class constraint system
into first class by introducing new fields
in the extended phase space.
As results, we obtain simultaneously the St\"uckelberg scalar term,
which is needed to cancel the gauge anomaly due to the mass term, and
the new type of Wess-Zumino action,
which is irrelevant to the gauge symmetry.
We also investigate the infrared property of the
Chern-Simons-Proca theory by using the Batalin-Tyutin formalism
comparing with the symplectic formalism.
As a result, we observe that the resulting theory is precisely
the gauge invariant Chern-Simons-Proca quantum mechanical
version of this theory.

\vspace{3cm}

PACS number : 11.10.Ef, 11.30.Ly, 11.15.Tk
\newpage

\begin{center}
\section{\bf Introduction}
\end{center}

The Dirac method has been widely used in the Hamiltonian formalism [1]
to quantize second class constraint systems.
However, since the resulting Dirac brackets are generally field-dependent
and nonlocal, and have a serious ordering problem
between field operators, these are under unfavorable circumstances
in finding canonically conjugate pairs.
On the other hand,
the quantization of first class  constraint systems [2,3] has been well
appreciated in a gauge invariant manner preserving
Becci-Rouet-Stora-Tyutin (BRST) symmetry [4,5].
This formalism has been extensively studied by
Batalin, Fradkin, and Tyutin [6,7] in canonical formalism,
and applied to various models [8-10]
obtaining the Wess-Zumino (WZ) action [11,12].
Recently, Banerjee [13] has applied the
Batalin-Tyutin (BT) Hamiltonian method [7] to
the abelian Chern-Simons (CS) field theory [14-16].
As a result, he has obtained the new type of an abelian WZ action,
which cannot be obtained in the usual path-integral framework.
Very recently, we have quantized the nonabelian case
by generalizing this BT formalism [17].
As shown in these works,
the nature of second class constraint algebra in the original theories
originates from the symplectic structure of CS term,
not due to the local gauge symmetry breaking.
Banerjee, Ghosh, and Banerjee [18] have also considered
a massive Maxwell theory.
As a result, the extra field in this approach has identified
with the St\"uckelberg scalar.
We have also quantized the abelian self-dual massive theory by
using this formalism, which interestingly produces both
the St\"uckelberg scalar and the new type of WZ [19].
There are some other interesting examples in this approach [20].

On the other hand, three-dimensional Chern-Simons gauge theories have
been attracting much attention because these play an important
role in the present development of the quantum Hall effect [21]
and the string theory [22].
The quantum mechanical version of the CS field theory has been studied
by Jackiw and collaborators through the phase space reductive
limiting procedure [23].
Recently, Baxter [24] has described a simple (2+1)-dimensional system
which allow in principle as the experimental verification of the CS
feature by introducing the R\"ontgen energy term [25].

In the present paper, we shall apply the BT Hamiltonian method [7]
to the Chern-Simons-Proca (CSP) theory [26]
revealing both the St\"uckelberg effect [27]
and the CS effect [13,17].
In Sec. 2, we apply the BT formalism to the CSP theory
in three dimensions which is gauge non-invariant.
By identifying the new fields $\rho$ and $\lambda$ with the
St\"uckelberg scalar and the WZ scalar, respectively,
we obtain simultaneously the St\"uckelberg scalar term related
to the explicit gauge-symmetry-breaking mass term
and the new type of WZ action
related to the symplectic structure of the CS term.
In Sec. 3, we also investigate
the quantum mechanical version of the CSP theory
by using the BT formalism comparing with the symplectic formalism [29],
which is the improved version of the Dirac method, and, in particular,
very effective for the first-order Lagrangian.
As a result, we observe that the resulting theory
is just the gauge invariant CSP quantum mechanical model.

\vspace{1cm}
\begin{center}
\section {\bf The Chern-Simons-Proca Theory}
\end{center}

We consider the abelian CSP model [26]
\begin{equation}
 S_{CSP}  =  \int d^3 x~ [
             - \frac{1}{2} {\kappa} \epsilon_{\mu\nu\rho}
             A^{\mu} \partial^{\nu} A^{\rho}
             + \frac{1}{2} m^2 A^\mu A_\mu
             ]
\end{equation}
by using the BT formalism.
Note that this action has an explicit mass term,
which breaks the gauge symmetry as the case of the Proca model [18],
and also the CS term, which has a different origin of the second class
constraint system.
Consequently, this action represents the second class constraint system
combined with two effects,
which can be easily confirmed by the standard Dirac analysis [1].
There are three primary constraints,
\begin{eqnarray}
\Omega_0 &\equiv& \pi_0 \approx 0, \nonumber \\
\Omega_i &\equiv& \pi_i + \frac{1}{2} {\kappa} \epsilon_{ij} A^j \approx 0,
                 ~~~~~( i = 1, 2 ),
\end{eqnarray}
and one secondary constraint,
\begin{equation}
\omega_3 \equiv m^2 A^0 - {\kappa} \epsilon_{ij} \partial^i A^j \approx 0,
\end{equation}
which is obtained by conserving $\Omega_0$ with the total Hamiltonian,
\begin{equation}
H_T = H_c + \int d^2x [ u^0 \Omega_0 + u^i \Omega_i ],
\end{equation}
where $H_c$ is the canonical Hamiltonian,
\begin{equation}
H_c = \int d^2x \left[
                      {\kappa} \epsilon_{ij} A^0 \partial^i A^j
                      + \frac{1}{2} m^2 \{ (A^i)^2 - (A^0)^2 \}
                      \right],
\end{equation}
and  $u^0$, $u^i$ are Lagrange multipliers.
No further constraints are generated via this procedure.
We find that all constraints are fully second class.
In order to carry out the simple algebraic manipulation,
it is, however, essential to redefine $\omega_3$
by using $\Omega_i$ as follows
\begin{eqnarray}
\Omega_3 &\equiv& \omega_3 + \partial^i \Omega_i \nonumber\\
         &=&     \partial^i \pi_i
               - \frac{1}{2} {\kappa} \epsilon_{ij} \partial^i A^j
               + m^2 A^0,
\end{eqnarray}
although the redefined constraints $\Omega_\alpha (\alpha=0,1,2,3)$
are still completely second class.
Otherwise, one will have a complicated constraint algebra including the
derivative terms, which is difficult
to handle.
Then, the modified constraint algebra is given by
\begin{eqnarray}
\Delta_{\alpha\beta}(x,y) &\equiv&
                  \{ \Omega_{\alpha}(x), \Omega_{\beta}(y) \} \nonumber\\
               &=&
          \left( \begin{array}{cccc}
         0 &          0         &    0   &    - m^2    \\
         0 &          0         &    {\kappa}   &     0        \\
         0 &         -{\kappa}         &    0   &     0        \\
         m^2 &        0         &    0    &     0
           \end{array}
          \right)
  \delta^2(x-y) ,
\end{eqnarray}
which reveals the simple second class nature
of the constraints $\Omega_\alpha$.

In order to convert this system into first class,
the first objective is
to transform $\Omega_\alpha$ into the first class
by extending a phase space.
Following the BT approach [7], we introduce
new auxiliary fields $\Phi^\alpha$,
and assume that the Poisson algebra of the new fields is given by
\begin{equation}
   \{ \Phi^\alpha(x), \Phi^\beta(y) \} = \omega^{\alpha\beta}(x,y),
\end{equation}
where $\omega^{\alpha\beta}$ is an antisymmetric matrix.
Then, the modified constraint in the extended phase space is given by
\begin{equation}
  \tilde{\Omega}_\alpha (\pi_\mu, A^\mu, \Phi^\beta)
         =  \Omega_\alpha + \sum_{n=1}^{\infty} \Omega_\alpha^{(n)};
                       ~~~~~~\Omega_\alpha^{(n)} \sim (\Phi^\beta)^n,
\end{equation}
satisfying the boundary condition,
$\tilde{\Omega}_\alpha(\pi_\mu, A^\mu, 0) = \Omega_\alpha$.
The first order correction term in the infinite series [7] is given by
\begin{equation}
  \Omega_\alpha^{(1)}(x) = \int d^2 y X_{\alpha\beta}(x,y)\Phi^\beta(y),
\end{equation}
and the first class  constraint algebra of $\tilde{\Omega}_\alpha$
requires the condition as follows
\begin{equation}
   \triangle_{\alpha\beta}(x,y) +
   \int d^2 w~ d^2 z~
        X_{\alpha\gamma}(x,w) \omega^{\gamma\delta}(w,z)
        X_{\delta\beta}(z,y) = 0.
\end{equation}
Among the solutions satisfying with the conditions (8) and (11),
we take a simple solution as follows
\begin{eqnarray}
\omega^{\alpha\beta} (x,y) &=&
    \left( \begin{array}{cccc}
         0 &          0         &    0   &    1    \\
         0 &          0         &    -1   &     0        \\
         0 &          1         &    0   &     0        \\
        -1 &        0         &    0    &     0
           \end{array}
   \right)
  \delta^2(x-y), \nonumber \\
X_{\alpha\beta} (x,y) &=&
      \left( \begin{array}{cccc}
         m &          0         &    0   &    0    \\
         0 &          \sqrt{\kappa}         &    0   &     0        \\
         0 &          0         &    \sqrt{\kappa}   &     0        \\
         0 &          0         &    0    &     m
           \end{array}
   \right)
  \delta^2(x-y).
\end{eqnarray}
There is an arbitrariness in choosing $\omega^{\alpha\beta}$,
which would naturally be manifested in Eq. (12). This just corresponds
to the canonical transformations in the extended phase space.
However, as has been shown in other calculations [13,17],
this choice of Eq. (12) gives the remarkable algebraic simplification.

Using Eqs. (9) and (12), the new set of constraints is found to be
\begin{eqnarray}
\tilde{\Omega}_0 &=& \Omega_0 + m \Phi^0,  \nonumber\\
\tilde{\Omega}_i &=& \Omega_i
                     + \sqrt{\kappa} \Phi^i, ~~~~~~~~(i=1,2), \nonumber\\
\tilde{\Omega}_3 &=& \Omega_3 + m \Phi^3,
\end{eqnarray}
which are strongly involutive,
\begin{equation}
\{ \tilde{\Omega}_{\alpha}, \tilde{\Omega}_{\beta} \} = 0.
\end{equation}
As a result, we have fully first class constraints
in the extended phase space
by applying the BT formalism systematically.
Observe further that only the $\Omega_\alpha^{(1)}$ contributes
in the series
(9) defining the first class constraint.
All higher order terms given by Eq. (9) vanish
as a consequence of the choice Eq. (12).
Recall the $\Phi^\alpha$ are the new variables satisfying the algebra
(8) with $\omega^{\alpha\beta}$  given  by Eq. (12).

The next step is to obtain the involutive Hamiltonian,
which naturally generates the secondary Gauss constraint
in the extended phase space.
It is given by the infinite series [7],
\begin{equation}
 \tilde{H} = H_c + \sum_{n=1}^{\infty} H^{(n)};
            ~~~~~H^{(n)} \sim (\Phi^\alpha)^n,
\end{equation}
satisfying the initial condition,
$\tilde{H}(\pi_\mu, A^\mu, 0) = H_c$.
The general solution [7] for the involution of $\tilde{H}$ is given by
\begin{equation}
  H^{(n)} = -\frac{1}{n} \int d^2 x d^2 y d^2 z~
              \Phi^\alpha(x) \omega_{\alpha\beta}(x,y)
              X^{\beta\gamma}(y,z) G_\gamma^{(n-1)}(z),
              ~~~(n \geq 1),
\end{equation}
where the generating functionals $G_\alpha^{(n)}$ are given by
\begin{eqnarray}
  G_\alpha^{(0)} &=& \{ \Omega_\alpha^{(0)}, H_c \},  \nonumber  \\
  G_\alpha^{(n)} &=& \{ \Omega_\alpha^{(0)}, H^{(n)} \}_{\cal O}
                    + \{ \Omega_\alpha^{(1)}, H^{(n-1)} \}_{\cal O}
                                       ~~~ (n \geq 1),
\end{eqnarray}
where the symbol ${\cal O}$ in Eq. (17) represents
that the Poisson brackets are calculated among the original variables,
{\it i.e.}, ${\cal O}=(\pi_\mu, A^\mu)$.
Here, $\omega_{\alpha\beta}$ and $X^{\alpha\beta}$ are
the inverse matrices of $\omega^{\alpha\beta}$ and
$X_{\alpha\beta}$ respectively.
Explicit calculations of $G_\alpha^{(0)}$ yield,
\begin{eqnarray}
G_0^{(0)} &=& m^2 A^0 - {\kappa} \epsilon_{ij} \partial^i A^j,
              \nonumber \\
G_i^{(0)} &=& - m^2 A^i - {\kappa} \epsilon_{ij} \partial^j A^0,
              \nonumber \\
G_3^{(0)} &=& m^2 \partial_i A^i,
\end{eqnarray}
which are substituted in Eq. (16) to obtain $H^{(1)}$,
\begin{equation}
H^{(1)} = \int d^2x  \left[
        m \Phi^0 \partial_i A^i
        + \frac{m^2}{\sqrt{\kappa}} \epsilon_{ij} \Phi^i A^j
        + \sqrt{\kappa} \Phi^i \partial_i A^0
        - \Phi^3 ( m A^0 - \frac{\kappa}{m} \epsilon_{ij} \partial^i A^j )
               \right].
\end{equation}
This is inserted back in Eq. (17) to deduce $G_\alpha^{(1)}$ as follows
\begin{eqnarray}
G_0^{(1)} &=& {\sqrt{\kappa}} \partial_i \Phi^i  + m \Phi^3, \nonumber \\
G_i^{(1)} &=& m \partial_i \Phi^0 + \frac{m^2}{\sqrt{\kappa}} \epsilon_{ij}
             \Phi^j - \frac{\kappa}{m} \epsilon_{ij} \partial^j \Phi^3,
             \nonumber\\
G_3^{(1)} &=& m \partial_i \partial^i \Phi^0 + \frac{m^2}{\sqrt{\kappa}}
             \epsilon_{ij} \partial^i \Phi^j,
\end{eqnarray}
which then yield $H^{(2)}$ from Eq. (16),
\begin{equation}
H^{(2)} = \int d^2x  \left[
       - \frac{1}{2} \partial_i \Phi^0 \partial^i \Phi^0
       + \frac{m}{\sqrt{\kappa}} \Phi^0 \epsilon_{ij} \partial^i \Phi^j
       + \frac{m^2}{2\kappa} \Phi^i \Phi^i
      - ( \frac{\sqrt{\kappa}}{m} \partial_i \Phi^i
          + \frac{1}{2} \Phi^3 ) \Phi^3
       \right].
\end{equation}
Since $G_\alpha^{(n)} = 0 ~~(n \geq 2)$, the final expression for the
involutive Hamiltonian after the $n=2$ finite truncations
is given by
\begin{equation}
\tilde H = H_c + H^{(1)} + H^{(2)},
\end{equation}
which is strongly involutive,
\begin{equation}
\{\tilde{\Omega}_\alpha, \tilde H\} = 0.
\end{equation}
According to the usual BT formalism,
this formally completes the operatorial conversion of the original
second class system with Hamiltonian $H_c$ and constraints
$\Omega_\alpha$ into the first class
with Hamiltonian $\tilde H$ and constraints $\tilde{\Omega}_\alpha$.

However, before performing the momentum integrations to obtain the
partition function in the configuration space,
it seems appropriate to comment on the strongly involutive Hamiltonian.
If we directly use this Hamiltonian,
we can not naturally generate
the first class Gauss' law constraint $\tilde{\Omega}_3$ from
the time evolution of the primary constraint $\tilde{\Omega}_0$,
which is the first class.
Therefore, in order to avoid this problem,
we use the equivalent first class Hamiltonian
without any loss of generality,
which only differs from the involutive Hamiltonian (22)
by adding a term proportional to the first class constraint
$\tilde{\Omega}_3$ as follows
\begin{equation}
\tilde{H}^{'} = \tilde{H} + \frac{1}{m} \Phi^3 \tilde{\Omega}_3.
\end{equation}
Then, this desired Hamiltonian $\tilde {H}^{'}$ automatically generates
the Gauss' law constraint
such that $\{ \tilde{\Omega}_0, \tilde{H}^{'} \} = \tilde{\Omega}_3$.
Note that when we act this modified Hamiltonian on physical states,
the difference with $\tilde{H}$ is trivial because such states are
annihilated by the first class constraint.
Similarly, the equations of motion for observable ({\it i.e.}
gauge invariant variables) will also be unaffected by this difference
since $\tilde{\Omega}_3$ can be regarded as
the generator of the gauge transformations.

We now derive the Lagrangian, which will include
both the St\"uckelberg effect and the CS effect, corresponding to
the Hamiltonian (24).
The first step is to identify the new variables
$\Phi^\alpha$ as canonically conjugate pairs in the Hamiltonian formalism
as follows
\begin{equation}
\Phi^\alpha \equiv ( m \rho,
                   \frac{1}{\sqrt{\kappa}} \pi_{\lambda},
                    \sqrt{\kappa} \lambda,
                    \frac{1}{m} \pi_{\rho})
\end{equation}
satisfying Eqs. (8) and (12).
The starting phase space partition function is then given
by the Faddeev formula [28],
\begin{equation}
Z =  \int  {\cal D} A^\mu {\cal D} \pi_\mu
           {\cal D} \lambda {\cal D} \pi_\lambda
           {\cal D} \rho {\cal D} \pi_\rho
          \prod_{\alpha,\beta = 0}^{3} \delta(\tilde{\Omega}_\alpha)
                               \delta(\Gamma_\beta)
          det \mid \{\tilde{\Omega}_\alpha,\Gamma_\beta\} \mid
          e^{iS'},
\end{equation}
where
\begin{equation}
S'  =  \int d^3x \left( \pi_\mu {\dot A}^\mu + \pi_\lambda {\dot \lambda}
                 + \pi_\rho {\dot \rho}
                 - \tilde{\cal H'}
           \right)
\end{equation}
with the Hamiltonian density $\tilde{\cal H}'$ corresponding
to $\tilde H'$, which is now expressed in
terms of  $\{ \rho, \pi_{\rho}, \lambda, \pi_\lambda \}$
instead of $\Phi^\alpha$.
The gauge fixing conditions $\Gamma_\alpha$ may be assumed
to be independent of the momenta so that these are considered
as the Faddeev-Popov type gauge conditions [28].

Next, we perform the momentum integrations to obtain the
configuration space partition function.
The  $\pi_0$, $\pi_1$, and $\pi_2$ integrations are
trivially performed by exploiting the delta functions
$~~\delta(\tilde{\Omega}_0)~ =~ \delta(\pi_0 + m^2 \rho)$,
$~~\delta(\tilde{\Omega}_1)~ =~
\delta(\pi_1+\frac{\kappa}{2}A^2+\pi_\lambda)$,
and
$~~\delta(\tilde{\Omega}_2)~ =~
\delta(\pi_2-\frac{\kappa}{2}A^1+{\kappa}\lambda)$,
respectively.
After exponentiating the remaining delta function
$\delta(\tilde{\Omega}_3) = \delta(-{\kappa}\epsilon_{ij}\partial^iA^j
+\partial_1\pi_\lambda+{\kappa}\partial_2\lambda+m^2A^0+\pi_\rho)$
with Fourier variable $\xi$ as
$\delta(\tilde{\Omega}_3)=\int{\cal D}\xi e^{-i\int d^3x
\xi\tilde{\Omega}_3}$
and transforming $A^0 \to A^0 + \xi$,
we obtain the action as follows
\begin{eqnarray}
S &=& \int d^3x~ \{
 - \frac{1}{2} \kappa \epsilon_{\mu\nu\rho} A^\mu \partial^\nu A^\rho
 + \frac{1}{2} m^2 A^\mu A_\mu \nonumber\\
  &+& \rho [ -m^2 (\dot{A^0} + \dot{\xi}) - m^2 \partial_i A^i
    - \frac{1}{2} m^2 \partial_i \partial^i \rho
    + m^2 \partial_1 \lambda - \frac{m^2}{\kappa} \partial_2 \pi_\lambda ]
            \nonumber\\
  &+& \pi_\rho [ \dot{\rho} - \frac{1}{2m^2} \pi_\rho - \xi ]
      + \lambda [ - {\kappa} \dot{A}^2 + m^2 A^1 - \kappa \partial_2 A^0
      - \frac{1}{2} m^2 \lambda ] \nonumber\\
     &+&  \pi_\lambda [ \dot{\lambda} - \dot{A}^1
        - \frac{m^2}{\kappa} A^2 + \partial^1 A^0
        - \frac{1}{2} \pi_\lambda ] - \frac{1}{2} m^2 \xi^2  \},
\end{eqnarray}
where the overdot means the time derivative,
and the corresponding measure is given by
\begin{equation}
[{\cal D} \mu] = {\cal D} A^\mu
                 {\cal D} \lambda
                 {\cal D} \pi_\lambda
                 {\cal D} \rho
                 {\cal D} \pi_\rho
                 {\cal D} \xi
                 \prod^3_{\beta = 0}
                \delta(\Gamma_{\beta}[A^0 + \xi, A^i, \lambda, \rho])
                det \mid \{\tilde{\Omega}_{\alpha}, \Gamma_{\beta}\} \mid,
\end{equation}
where
$A^0 \to A^0 + \xi$ transformation is naturally understood
in the gauge fixing condition $\Gamma_{\beta}$.

Note that the original theory is easily reproduced in one line,
{\it i.e.}, if we choose the unitary gauge
\begin{equation}
\Gamma_\alpha = ( \rho, \pi_\lambda, \lambda, \pi_\rho ),
\end{equation}
and integrate over $\xi$.
Then, one can easily realize that
the new fields $\Phi^\alpha$ are nothing but the gauge degrees of
freedom, which can be removed by utilizing the gauge symmetry.

Now, we perform the Gaussian integration over $\pi_\rho$.
Then all terms including $\xi$ in the action are canceled out,
and integrating over $\pi_\lambda$
the resultant action is finally obtained as follows
\begin{eqnarray}
S &=& S_{St} + S_{NWZ} + S_{B}~~;  \nonumber \\
S_{St} &=&
           \int d^3x
        \{   - \frac{1}{2} \kappa \epsilon_{\mu\nu\rho}
                A^{\mu} \partial^{\nu} A^{\rho}
             + \frac{1}{2} m^2 (A_\mu + \partial_\mu \rho )^2
                          \},                              \nonumber \\
S_{NWZ}  &=& \int d^3x
         \{ \frac{\kappa^2}{2m^2} [ {\dot \lambda} + F_{01}
                         + \frac{m^2}{\kappa}
                         (A_2 + \partial_2 \rho) ]^2       \nonumber \\
       &&~~~~~~~~~+~ \kappa \lambda [ F_{02}
                      - \frac{m^2}{\kappa} (A_1 + \partial_1 \rho)
                      - \frac{m^2}{2\kappa} \lambda ]    \}, \nonumber \\
S_{B} &=& - \int d^3x~ \partial_\mu (m^2 \rho A^\mu )
\end{eqnarray}
where $F_{\mu \nu}=\partial_\mu A_\nu - \partial_\nu A_\mu$.
Note that $S_{St}$ is an expected St\"uckelberg scalar term,
which is needed to cancel the gauge anomaly due to the explicit
gauge-symmetry-breaking mass term [18],
$S_{NWZ}$ is the new type of WZ term
due to the symplectic structure of the CS term [13,19],
which is irrelevant to the gauge symmetry,
and $S_B$ is the boundary term, which is also needed to make the
second class system into the first class.
The corresponding Liouville measure just comprises
the configuration space variables as follows
\begin{equation}
[{\cal D} \mu] = {\cal D} A^\mu
                 {\cal D} \lambda
                 {\cal D} \rho
                 {\cal D} \xi
                 \prod^3_{\beta = 0}
                \delta(\Gamma_{\beta}[A^0 + \xi, A^i, \lambda, \rho])
                det \mid \{\tilde{\Omega}_{\alpha}, \Gamma_{\beta}\} \mid.
\end{equation}
This action $S$ is invariant up to the total divergence
under the gauge transformations as
$\delta A_\mu = \partial_\mu \Lambda$, $\delta \rho = - \Lambda$, and
$\delta \lambda = 0$.

Note that starting from the action (31) with the boundary term $S_{B}$,
we can easily reproduce the same set of all first class constraints
$\tilde{\Omega}_\alpha$, and the Hamiltonian such that
\begin{eqnarray}
H = H_c &+& \int d^2x~~ [ \pi_\lambda \partial_1 A_0
                         - \frac{m^2}{\kappa} \pi_\lambda A_2
      + m^2 \rho \partial_i A^i + \kappa \lambda \partial_2 A_0
      + m^2 \lambda A_1         \nonumber \\
  &+& \frac{m^2}{2\kappa^2} \pi_\lambda^2
      - \frac{m^2}{\kappa} \pi_\lambda \partial_2 \rho
      - \frac{1}{2} m^2 \partial_i \rho \partial^i \rho
      + m^2 \lambda \partial_1 \rho + \frac{1}{2} m^2 \lambda^2
      + \frac{1}{2m^2} \pi_\rho^2].
\end{eqnarray}
Then, if we add a term proportional to the constraint
$\tilde{\Omega}_3$, $i.e., \frac{1}{m^2} \pi_\rho \tilde{\Omega}_3$,
which is trivial when acting on the physical Hilbert space,
to the above Hamiltonian (33),
we can obtain the original involutive Hamiltonian (22).
Furthermore, this difference is also trivial in the construction
of the functional integral because the constraint $\tilde{\Omega}_3$
is strongly implemented by the delta function
$\delta(\tilde{\Omega}_3)$ in Eq. (26).
Therefore, we have shown that the constraints and the Hamiltonian
following from the action (31) are effectively equivalent to
the original Hamiltonian embedding structure.
As results, through the BT quantization procedure,
we have found that
the St\"uckelberg scalar $\rho$ is naturally introduced in the mass term,
and this $\rho$ as well as the WZ scalar $\lambda$ is also included
in the new type of WZ action.

We also note that
if we ignore the boundary term $S_{B}$ in this action,
we cannot directly obtain the involutive first class Hamiltonian
as the case of the Proca theory explained in Ref. [18]
because this boundary term plays the important role in this procedure.

Finally, note that in the trivial limit $\kappa \rightarrow m$,
the action (31) exactly reduces to the self-dual massive theory
having all the first class constraints, which has
recently been derived in Ref. [19].
The limit $m \rightarrow 0$ is non-trivial because
the action has the $m^{-2}$ term.
In fact, we can easily find that the auxiliary field $\pi_\lambda$ is not
well-defined in the action for the case of this limit
because the $\pi_\lambda$ contains the $m^{-2}$ term, {\it i.e.},
$\pi_\lambda = \frac{\kappa^2}{m^2}(\dot{\lambda}+F_{01})
+\kappa(A_2+\partial_2 \rho)$.
Therefore, we have to pay attention to when the momentum integrations
are performed in the below of Eq. (28).
Avoiding this situation, if we simultaneously take the limit
$(\dot{\lambda}+F_{01}) \rightarrow 0$ with
the $m \rightarrow 0$,
we obtain the delta function $\delta(\dot{\lambda}+F_{01})$
in the measure part
when we perform the momentum integration over $\pi_\lambda$
resulting to the case of the pure CS action.
Then, one can finally find that the CSP theory exactly reduces to
the pure CS case having all the first class constraints [19].

\begin{center}
\section{\bf The Chern-Simons-Proca Quantum Mechanics}
\end{center}

\begin{center}
\subsection{\bf The Symplectic Quantization of the CSP Quantum Mechanics}
\end{center}

Let us briefly discuss the symplectic quantization,
which is very effective for the first-order Lagrangian [29],
of the CSP quantum mechanics.
We start the following infrared limit action [23],
which is already first-order, of the CSP theory
with the Coulomb gauge
\begin{equation}
 S_o =  \int dt L_o =
      \int dt \left[ \frac{1}{2} {\kappa} \epsilon_{ij} q^i \dot{q}^j
      - \frac{1}{2} m q^i q^i \right],
\end{equation}
where $\epsilon_{12}=\epsilon^{12}=1$.
Following the symplectic formalism, we first rewrite the action as follows
\begin{equation}
 S_o = \int dt \left[ - \frac{\kappa}{2} q^2 \dot{q}^1
                + \frac{\kappa}{2} q^1 \dot{q}^2 - H^{(0)} \right],
\end{equation}
where $H^{(0)}=\frac{m}{2} [ (q^1)^2 + (q^2)^2 ]$ is the usual canonical
Hamiltonian and the superscript denotes the number of iterations [29].
Note that since the action (35) is already first-ordered from the start,
we do not need to introduce auxiliary fields such as conjugate momenta.
Then, we set symplectic variables
$\xi^{(0)i}=(q^1, q^2)$, and symplectically conjugated momenta
$a_i^{(0)}=(-\frac{\kappa}{2}q^2, \frac{\kappa}{2}q^1)$.

Now, symplectic 2-form matrix $f_{ij}$,
which consists of the essential part for finding the generalized brackets
of the symplectic formalism, is obtained as follows
\begin{equation}
f^{(0)}_{ij} = \frac{\partial a^{(0)}_j}{\partial \xi^{(0)i}}
           - \frac{\partial a^{(0)}_i}{\partial \xi^{(0)j}}
       =  \kappa \left( \begin{array}{cc}
                      0 &       1    \\
                     -1 &       0     \\
           \end{array} \right)
\end{equation}
having the antisymmetric property.
This matrix is not singular, and thus has the inverse as follows
\begin{eqnarray}
f^{(0)ij} =
               - \frac{1}{\kappa}
               \left( \begin{array}{cc}
                      0 &     1    \\
                     -1 &     0     \\
                       \end{array} \right),
\end{eqnarray}
which gives the generalized symplectic brackets
\begin{equation}
\{q^i, q^j \} = - \frac{1}{\kappa}\epsilon^{ij}.
\end{equation}
These are exactly same as the Dirac brackets [23]
when we analyze the system
through the usual Dirac's method.

It is appropriate to comment on the symplectic formalism [29].
In general, the symplectic 2-form matrix
is not invertible at the first stage of iterations.
Then, we can find some zero modes, which are related to generate
the constraints in the symplectic formulation,
and incorporate them into the canonical sector
with some auxiliary variables to find the nonvanishing symplectic
2-form matrix.
If we can find the invertible symplectic 2-form matrix
at the finite stage of iterations, the inverse of the matrix gives
the generalized brackets,
which are equivalent to the usual Dirac brackets.
However, when we can not find the invertible matrix even at the infinite
stages of iterations,
we can say the system has a gauge symmetry and
use the zero modes to obtain the concrete rules of transformations [29].
Especially in the case of the CSP quantum mechanical model,
we have the symplectic 2-form matrix at the first stage of iterations.
Thus the system has no constraints
in the symplectic quantization formalism,
while this system has second class constraints [23]
in the standard Dirac formalism as follows
\begin{equation}
\Omega_i \equiv p_i + \frac{1}{2} \kappa \epsilon_{ij} q^j \approx 0
{}~~~~~(i=1,2),
\end{equation}
where $p_i=\frac{\partial L_o}{\partial \dot{q}^i}=-\frac{1}{2}
\epsilon_{ij}q^j$ are canonical momenta.

Now, using the generalized brackets, we can easily find
the Hamilton equations as follows
\begin{equation}
\dot{q}^i = \{q^i, H^{(0)} \} = -\frac{m}{\kappa}\epsilon^{ij} q^j.
\end{equation}
These equations give a single simple harmonic oscillator
\begin{equation}
\dot{\gamma}(t) \equiv \dot{q}^1 + i \dot{q}^2 = i\omega \gamma(t),
\end{equation}
with the frequency $\omega=\frac{m}{\kappa}$ as usual.
In the next section, we will show that starting the gauge non-invariant
action (34), we can obtain the gauge invariant version describing
the simple harmonic oscillator by the BT formalism.

\begin{center}
\subsection{\bf The BT Quantization of the CSP Quantum Mechanics}
\end{center}

Now let us analyze the action (34) in the BT quantization
as in the previous Section 2.
The first observed fact through the usual Dirac's procedure [1] is that
the action represents a second class constraint system, $i.e.,$
there are two primary constraints (39),
and no further constraints are generated through the time evolution
of these constraints with the total Hamiltonian,
\begin{equation}
H_T = H_c + u^i \Omega_i ,
\end{equation}
where $H_c$ is the canonical Hamiltonian,
\begin{equation}
H_c = p_i \dot{q}^i - L_o = \frac{m}{2} q^i q^i,
\end{equation}
and $u^i$ are Lagrange multipliers.
Then, the constraint algebra is given by
\begin{eqnarray}
\Delta_{ij} &\equiv&
                  \{ \Omega_i, \Omega_j \} = \kappa \epsilon_{ij}
                  \nonumber\\
               &=&
               \kappa
          \left( \begin{array}{cc}
         0 &          1 \\
         -1 &  0       \\
           \end{array} \right),
\end{eqnarray}
which reveals the simple second class nature
of the constraints $\Omega_i$.

In order to convert this system into first class,
the first objective is
to transform $\Omega_i$ into the first class
by extending the phase space.
Following the BT approach [7], we introduce the matrix (8) as follows
\begin{equation}
\omega^{ij} =
    \left( \begin{array}{cc}
         0 &          -1    \\
         1 &          0      \\
           \end{array}
   \right) .
\end{equation}
Then the other matrix $X_{ij}$ in Eq. (10) is obtained by
solving Eq. (11) with the $\Delta_{ij}$ given by Eq. (45),
\begin{equation}
X_{ij} =  \sqrt{\kappa}
      \left( \begin{array}{cc}
         1 &    0   \\
         0 &    1   \\
           \end{array}
   \right) .
\end{equation}
There is an arbitrariness in choosing $\omega^{ij}$,
which would naturally be manifested in Eq. (45)
as explained in the Section 2.

Using Eqs. (9), (45) and (46), the new set of constraints is found to be
\begin{equation}
\tilde{\Omega}_i = \Omega_i + \sqrt{\kappa} \Phi^i,
\end{equation}
which are strongly involutive,
\begin{equation}
\{ \tilde{\Omega}_i, \tilde{\Omega}_j \} = 0.
\end{equation}

The next step is to obtain the involutive Hamiltonian.
The generating functions $G_i^{(n)}$ are obtained from Eq. (17).
It is noteworthy that there are only two terms $\Omega_i$ and
$\Omega_i^{(1)}$ in the expansion (47)
due to the choice (45) and (46).
Explicit calculations yield,
\begin{equation}
G_i^{(0)} = - m q^i ,
\end{equation}
which are substituted in Eq. (16) to obtain $H^{(1)}$,
\begin{equation}
H^{(1)} =   \frac{m}{\sqrt{\kappa}} \epsilon_{ij} \Phi^i q^j.
\end{equation}
This is inserted back in Eq. (17) to deduce $G_i^{(1)}$ as follows
\begin{equation}
G_i^{(1)} = \frac{m}{\sqrt{\kappa}} \epsilon_{ij} \Phi^j ,
\end{equation}
which then yield $H^{(2)}$ from Eq. (16),
\begin{equation}
H^{(2)} = \frac{m}{2\kappa} \Phi^i \Phi^i.
\end{equation}
Since $G_i^{(n)} = 0 ~~(n \geq 2)$, the final expression for the
involutive Hamiltonian after the $n=2$ finite truncations
is given by
\begin{equation}
\tilde H = H_c + H^{(1)} + H^{(2)}
         = \frac{m}{2} [ q^i q^i
          + \frac{2}{\sqrt{\kappa}} \epsilon_{ij} \Phi^i q^j
          + \frac{1}{\kappa} \Phi^i \Phi^i ].
\end{equation}
which, by construction, is strongly involutive,
\begin{equation}
\{\tilde{\Omega}_i, \tilde H\} = 0.
\end{equation}
This completes the operatorial conversion of the original
second class system with Hamiltonian $H_c$ and constraints
$\Omega_i$ into the first class
with Hamiltonian $\tilde H$ and constraints $\tilde{\Omega}_i$.

We now derive the gauge invariant Lagrangian corresponding
to the Hamiltonian (53).
The first step is to identify the new variables
$\Phi^i$ as canonically conjugate pairs in the Hamiltonian formalism,
\begin{equation}
\Phi^\alpha \equiv ( \frac{1}{\sqrt{\kappa}} P_X, \sqrt{\kappa} X),
\end{equation}
satisfying Eqs. (8), (45) and (46).
The starting phase space partition function is then given
by the Faddeev formula,
\begin{equation}
Z =  \int  {\cal D} q^i {\cal D} p_i
           {\cal D} P_X {\cal D} X
          \prod_{i,j} \delta(\tilde{\Omega}_i) \delta(\Gamma_j)
          det \mid \{\tilde{\Omega}_i, \Gamma_j \} \mid
          e^{iS},
\end{equation}
where
\begin{equation}
S  =  \int dt \left( p_i {\dot q}^i + P_X {\dot X} - \tilde{H}
           \right).
\end{equation}
As similar to the Section 2, the gauge fixing
conditions $\Gamma_i$ may be assumed to be independent
of the momenta so that these are considered as the Faddeev-Popov type
gauge conditions.

Next, we perform the momentum integrations to obtain the
configuration space partition function.
The  $p_1$, and $p_2$ integrations are
trivially performed by exploiting the delta functions
$~~\delta(\tilde{\Omega}_1)~
=~ \delta(p_1+\frac{\kappa}{2}q^2+ P_X) $,
and
$~~\delta(\tilde{\Omega}_2)
{}~ =~ \delta(p_2-\frac{\kappa}{2}q^1+ \kappa X) $,
respectively.
Then, we obtain the action as follows
\begin{eqnarray}
S &=& \int dt~ \{ \frac{1}{2} \kappa \epsilon_{ij} q^i {\dot q}^j
                 - \frac{1}{2} m q^i q^i  \nonumber\\
  &+&
       X ( -\kappa \dot{q}^2 + m q^1 - \frac{1}{2} m X)
      + P_X ( -\dot{q}^1 + \dot{X} - \frac{m}{\kappa} q^2
             - \frac{m}{2\kappa^2} P_X ) \} ,
\end{eqnarray}
and the corresponding measure is given by
\begin{equation}
[{\cal D} \mu] = {\cal D} q^i {\cal D} X {\cal D} P_X
          \prod_{i,j} \delta(\Gamma_j)
          det \mid \{\tilde{\Omega}_i, \Gamma_j \} \mid.
\end{equation}
Note that the original quantum mechanical model is also reproduced
when we choose the unitary gauge
such that $\Gamma_\alpha \equiv (X, P_X)$ as the case of the CSP theory.

Now, we perform the Gaussian integration over $P_X$.
Then, the resultant action is finally obtained as follows
\begin{eqnarray}
S &=& S_o + S_{NWZ} ~~~; \nonumber \\
S_{NWZ} &=& \int dt~\{
                     \frac{\kappa^2}{2m}(\dot{q}^1-\dot{X})^2
                     + \frac{m^2}{2}(q^2)^2
                     + \kappa q^2 \dot{q}^1
                     + m X q^1 - \frac{1}{2} m X^2
                     \}.
\end{eqnarray}
We can rewrite the above action with the boundary term
more compactly as follows
\begin{eqnarray}
S  &=& \int dt~
              \{ \frac{\kappa^2}{2m} (\dot{q}^1 - \dot{X})^2
              - \frac{m}{2} (q^1 - X)^2 \}
      + S_B~~; \nonumber \\
S_B &=&   \int dt~
               \frac{d}{dt} (-\kappa q^2 X + \frac{\kappa}{2}q^1 q^2 ).
\end{eqnarray}
The corresponding Liouville measure just comprises
the configuration space variables as follows
\begin{equation}
[{\cal D} \mu] = {\cal D} q^i {\cal D} X
          \prod_{i,j} \delta(\Gamma_j)
          det \mid \{\tilde{\Omega}_i, \Gamma_j \} \mid.
\end{equation}

The action (61) is invariant up to the total divergence
under the transformation
$\delta q^1 = \epsilon(t)$ and $\delta X = \epsilon(t)$,
which are just the gauge transformations
of the CSP quantum mechanical model.
Note that the $q^2$ variable has not appeared in Eq. (61)
except the boundary term $S_B$.
Furthermore, the action except the term $S_B$ is just a
usual harmonic oscillator having the frequency $\omega=\frac{m}{\kappa}$
as in the previous section
when we define a quantity such that $\gamma(t)=q^1-X$.
Since $\delta \gamma(t)$ is invariant under the above transformations,
it is a physical quantity.
As a result, starting from the gauge non-invariant system (34),
we obtain the gauge invariant version describing the harmonic
oscillator in the BT formalism.

Finally, we would like to comment that the gauge invariant action (61)
is not separated into the original action $S_o$
and the new type of the WZ action $S_{NWZ}$.
This is because $X$ is nothing but the gauge degree of freedom.

\begin{center}
\section{\bf Conclusion}
\end{center}

In conclusion, we have applied the Batalin-Tyutin method,
which converts the second class system into the first class,
to the CSP theory and the quantum mechanical version of this theory.
For the CSP case, we have shown that if we ignore the boundary term
in action (31),
the direct connection with the Lagrangian embedding of
St\"uckelberg scalar can be made by explicitly evaluating the momentum
integrals in the extended phase space partition function using
the Faddeev-Popov-like gauges,
and identifying the extra field $\rho$ introduced
in our Hamiltonian formalism with the conventional St\"uckelberg scalar
needed to cancel the gauge anomaly due to the mass term.
We have also obtained a new type of
WZ action $S_{NWZ}$ containing the WZ scalar $\lambda$,
which is irrelevant to the gauge anomaly.
Furthermore, we should also keep the boundary term $S_B$.
Otherwise, we cannot reproduce the original first class system.
Note that the St\"uckelberg scalar $\rho$ is also included in $S_{NWZ}$
in order to maintain the gauge invariance of the
$S_{NWZ}$ related to the CS effect in the action (31).

On the other hand, we have observed that
the infrared limit of the CSP theory
is precisely the gauge invariant CSP quantum mechanical model
by using the BT formalism comparing with the symplectic formalism.
Even though we can find a harmonic oscillator solution
by solving the Hamilton equations of motion
after applying the standard Dirac method,
we have applied the symplectic method since it is more intuitive
when we directly find the generalized brackets from the Hamilton equation
$f_{ij}{\dot \xi}^j = \frac{\partial}{\partial \xi^i}H(\xi)$
for the first-order system.
In other words, if $f_{ij}$ has an inverse,
the Hamilton equation is easily obtained through
${\dot \xi}^i = \{ \xi^i, ~H(\xi) \}
= \{ \xi^i,~ \xi^j \} \frac{\partial H(\xi)}{\partial \xi^j}$,
where $\{ \xi^i,~ \xi^j \}$ is the generalized bracket.
Furthermore, by applying the BT formalism,
we have firstly realized the harmonic oscillator at the action level,
which is manifestly gauge invariant in the CSP quantum mechanical
model.

\vspace{1cm}

\begin{center}
\section*{Acknowledgements}
\end{center}

We would like to thank Prof. K. Y. Kim and  Dr. S. -K. Kim
for useful comments and extensive discussions during the course of
this work.
The present study was supported by
the Basic Science Research Institute Program,
Ministry of Education, Project No. {\bf BSRI}-94-2414.

\end{document}